# Twisted Type-II Rashba Homobilayer: A Platform for Tunable Topological Flat Bands

[+]Xilong Xu,[1] [+]Haonan Wang,[1] Li Yang*,[1,2]

[1]Department of Physics, Washington University in St. Louis, St. Louis, Missouri 63130, USA

[2]Institute of Materials Science and Engineering, Washington University in St. Louis, St. Louis, Missouri 63130, USA

**Abstract**

The recent discovery of topological flat bands in twisted transition metal dichalcogenide homobilayers and multilayer graphene has sparked significant research interest. We propose a new platform for realizing tunable topological moiré flat bands: twisted type-II Rashba homobilayers. The interplay between Rashba spin-orbit coupling and interlayer interactions generates an effective pseudo-antiferromagnetic field, opening a gap within the Dirac cone with non-zero Berry curvature. Using twisted BiTeI bilayers as an example, we predict the emergence of flat topological bands with a remarkably narrow bandwidth (below 20 meV). Notably, the system undergoes a transition from a valley Hall insulator to a quantum spin Hall insulator as the twisting angle increases. This transition arises from a competition between the twisting-driven effective spin-orbit coupling and sublattice onsite energies presented in type-II Rashba moiré structures. The high tunability of Rashba materials in terms of the spin-orbit coupling strength, interlayer interaction, and twisting angle expands the range of materials suitable for realizing and manipulating correlated topological properties.

**Introduction**

The interplay between electronic topology and correlation gives rise to numerous exotic quantum phases, [1–6] such as the fractional quantum Hall (FQH) states in Landau levels. [7–12] Recently, fractional quantum anomalous Hall (QAH) effect in absence of Landau levels has been observed in moiré systems such as twisted bilayer $MoTe_2$ (t$MoTe_2$) and multilayer graphene/hexagonal boron nitride moiré superlattices, spurring tremendous research interest. [13–19] In these systems, topological moiré flat bands play an essential role in harboring correlation-driven fractional states. [20–22] For example, in rhombohedral-stacked $MoTe_2$ moiré bilayers, the Kane-Mele flat bands have been theoretically predicted, [23,24] and integer QAH states as well as quantum spin Hall (QSH) states have also been observed at integer filling factors. [17] Hence discovering new structures harboring topological flat bands are crucial for realizing and understanding topological physics in many-electron systems.

The Rashba effect, generally existing in the materials with strong spin-orbit coupling (SOC) and broken inversion symmetry, is capable of creating spin-polarized electron states in the absence of external magnetic fields. [25–28] This effect is characterized by a broken spin degeneracy, leading to quasi-particle bands with opposite spin textures that are momentum-offset [29,30] The remarkable tunability of the Rashba effect has enabled the observation of various well-known phenomena including the QAH effect, spin Hall effect, spin-charge conversion, and spin-torque in semiconductor devices. [25,29–31] Particularly, the Rashba-like bands are considered as a Dirac-cone surrounded by metallic states, indicating its promising potential for realizing topological states. However, its small effective mass and accompanying metallic state make it challenging to isolate those nontrivial bands and be measured via transport experiments to realize decisive topological properties.

In this study, we propose the twisted type-II Rashba structure as a promising platform for realizing and controlling flat bands with rich topological phases. Unlike conventional non-centrosymmetric Rashba materials, the type-II Rashba structure is centrosymmetric, which can be achieved by stacking two oppositely polarized layers. Using twisted BiTeI bilayer as an example, the combination of Rashba SOC and interlayer hopping creates a pseudo-antiferromagnetic field, which preserves time-reversal symmetry but induces an energy gap in the Dirac cone with nonvanishing Berry curvature. When the bilayer structure is further twisted, we find the topological $Z_2$ flat bands as well as QSH insulating states. Such QSH states feature the Kane-Mele model with an effective SOC induced by the moiré pseudo-Zeeman potential. A particularly intriguing observation is that the system undergoes a topological phase transition from the valley Hall (VH) phase to the QSH phase as the twist angle increases to 3.3°. This transition arises from the competition between the sublattice energy difference and effective SOC. Given the tunability of the Rashba intensity, interlayer interaction, and twist angle, our work highlights the potential of Rashba materials as a new family of material platforms for realizing and manipulating topological properties.

*Type-II Rashba effect:* Rashba effects are characterized by the two parabolic bands crossing over with spin polarizations in a helical texture, where the spin is oriented perpendicularly to the polar direction, as shown in the left panel of **Fig. 1(a)**. The Rashba effect widely exists in materials with broken inversion symmetry. Interestingly, a newly proposed Rashba effect can also exist in centrosymmetric materials by stacking two identical layers with the antiparallel polarization directions, which is referred to as the type-II Rashba effect (the right panel of **Fig. 1(a)**) [30,32–34] To date various type-II Rashba materials have been successfully fabricated, and they are considered as a key ingredient in the next generation of spin field-effect transistors. [32,35,36] To capture the features of a type-II Rashba bilayer structure, a minimized Hamiltonian can be written as (see Text S1 in Supplementary Information)

$$H(\boldsymbol{k}) = \frac{\hbar^2 k^2}{2m^*}\zeta_0 + \alpha_R \zeta_z (k_x s_y - k_y s_x) + \alpha_I \zeta_x s_0$$

$$= \begin{pmatrix} \frac{\hbar^2 k^2}{2m^*} + \alpha_R (k_x s_y - k_y s_x) & \alpha_I s_0 \\ \alpha_I s_0 & \frac{\hbar^2 k^2}{2m^*} - \alpha_R (k_x s_y - k_y s_x) \end{pmatrix}, \quad (1)$$

where $\zeta$ and $s$ are Pauli matrices that represent the layer pseudospins and spins, respectively. The off-diagonal block matrices stand for interlayer hopping. $m^*$ and $\alpha_R$ denote the effective mass and Rashba SOC strength, respectively, and $\alpha_I$ is the strength of interlayer hopping. Because the interlayer hopping conserves spin, this Hamiltonian preserves the time-reversal symmetry, satisfying $TH(\boldsymbol{k})T^{-1} = H(-\boldsymbol{k})$, where $T = i\zeta_0 s_y \mathcal{K}$ and $\mathcal{K}$ denotes the complex conjugation operator. Moreover, it also preserves the inversion symmetry $PH(\boldsymbol{k})P^{-1} = H(-\boldsymbol{k})$, where $P = \zeta_x s_0$. As a result, all bands exhibit double degeneracy in type-II Rashba materials, and the Berry curvature for the Kramers' degenerate pairs is vanishing. However, the Berry curvature for each band in the Kramers' pairs would remain nonzero. To see this, a unitary transformation, $U = \frac{1}{\sqrt{2}}(\zeta_0 s_0 - i\zeta_y s_z)$, can be performed on the Hamiltonian, leading

to

$$H_2(\mathbf{k}) = UH(\mathbf{k})U^\dagger = \frac{\hbar^2 k^2}{2m^*} + \zeta_z\big(\alpha_R(k_x s_y - k_y s_x) + \alpha_I s_z\big), \qquad (2)$$

where the second term essentially describes two two-dimensional (2D) massive Dirac fermions. At $\mathbf{k} = 0$, a band gap with $2\alpha_I$ is induced by the interlayer hopping term $\alpha_I \zeta_z s_z$ which can be regarded as a pseudo-antiferromagnetic field, as illustrated in right panel of **Fig. 1(a)**. Consequently, each band carries a nonvanishing Berry curvature with the Berry phase of $\pi$, and the in-plane Rashba spin texture around the Dirac cone is switched in the out-of-plane direction. Moreover, the Dirac-cone band gap increases as the interlayer hopping ($\alpha_I$) increases, as depicted in **Fig. 1(b)**. It is important to note that within a certain range of interlayer interaction strength, the bottom band becomes flat. This will help the formation of flat bands in twisted structures, as will be discussed in the following.

*Twisted type-II Rashba homobilayer:* The nonzero Berry curvature indicates the presence of topological states in type-II Rashba structures. However, these states are concealed within the Kramers' pairs due to the *PT* symmetry. Furthermore, the existence of parabolic bands results in metallic states within the Dirac-cone band gap, thereby impeding the probing of topological states. To break the combined *PT* symmetry and gap out these metallic states, a weak periodic potential can be achieved by twisting the type-II Rashba homobilayers. This motivates our search for vdW materials with strong Rashba effects and weak interlayer interactions.

BiTeI is a layered vdW material with an out-of-plane polarization and strong Rashba effect. [37–39] It has been fabricated and exfoliated [40,41] and exhibits a large bulk photovoltaic effect. [42,43] As shown in **Fig. 1(c)**, its monolayer structure consists of triple sublayers, similar to the *T*-phase of transition metal dichalcogenides (TMDs). The band structure of monolayer BiTeI is plotted in **Fig. 1(d)**. Notably, the bottom two conduction bands exhibit a typical Rashba band splitting around the $\Gamma$ point, while the top valence bands are more hybridized by different atomic orbitals that obscure the feature of Rashba splitting. Therefore, we mainly focus on the bottom two conduction bands around the $\Gamma$ point in the following. These two conduction bands of the monolayer can be described by $H = \frac{\hbar^2 k^2}{2m^*} + \alpha_R(k_x s_y - k_y s_x)$ with an effective mass $m^* = 0.49\, m_e$ and Rashba strength $\alpha_R = 1.53$ eV·Å, which is similar with those typical Rashba materials, such as GeTe (~2.5 eV·Å). [44]

To construct the type-II Rashba structure, two oppositely polarized BiTeI monolayers are stacked. Depending on the interfacial atoms, there are two basic types of interfaces: Te-Te and I-I. We will discuss the I-I interface in the main manuscript, which is shown in the inset of **Fig. 1(e)**. The Te-Te interface yields similar results, and the details are provided in the Supplementary Information. The band structure of type-II bilayer BiTeI is presented in **Fig. 1(e)**. As expected, the Dirac cone is now gapped due to interlayer hopping, forming a pair of parabolic-like energy dispersions. Each band is doubly degenerate due to the *PT* symmetry. This agrees well with our model Hamiltonian Eqs. (1) and (2) and **Fig. 1(b)**. As we twist the homobilayer BiTeI at a commensurate small twist angle to form moiré superlattices, the local interlayer atomic displacement will go through a slow periodic modulation that allows various local atomic stackings in the moiré unit cell. Because of the three-fold rotational symmetry, there are three high-symmetry stackings, i.e., AA, AB, and AC, in the moiré unit cell (**Fig. 1(f)**). All three high-symmetry stackings feature the similar type-II Rashba band structure but with different Dirac-cone gaps due to their different interlayer interactions. (See Table S1 in Supplementary Information)

To capture the moiré modulations on interlayer interactions, the effective moiré Hamiltonian of the four lowest-energy conduction bands can be written as: [21,23,24]

$$H_{Moiré}(\boldsymbol{r}, \boldsymbol{k}) = \left(\frac{\hbar^2 k^2}{2m^*} + \Delta(\boldsymbol{r})\right) + \zeta_z\big(\alpha_R(k_x s_y - k_y s_x) + \alpha_I(\boldsymbol{r}) s_z\big), \qquad (3)$$

where the interlayer hopping now becomes as a periodic function, and the term $\alpha_I(\boldsymbol{r})\zeta_z s_z$ contributes to a pair of opposite Zeeman moiré potentials. With the three-fold rotational symmetry, $\alpha_I(\boldsymbol{r})$ can be expanded as $\alpha_I(\boldsymbol{r}) = M_0 + M_1 \sum_{j=1,3,5} \cos(\boldsymbol{G}_j \cdot \boldsymbol{r} + \psi)$, where $\boldsymbol{G}_j = \frac{8\pi \sin\left(\frac{\theta}{2}\right)}{\sqrt{3} a_0} \left(\cos\left(\frac{j-1}{3}\right)\pi, \sin\left(\frac{j-1}{3}\right)\pi\right)$ denotes the three moiré reciprocal lattice vectors, $\theta$ and $a_0 = 3.8$ Å denote twist angle and lattice constant of the unit cell of BiTeI, respectively. Besides, $\Delta(\boldsymbol{r}) = V_0 + V_1 \sum_{j=1,3,5} \cos(\boldsymbol{G}_j \cdot \boldsymbol{r} + \phi)$ stands for the scalar moiré potential that originates from modulation on the conduction band minimum over the moiré period. All the parameters in $\alpha_I(\boldsymbol{r})$ and $\Delta(\boldsymbol{r})$ can be determined from first-principles simulations of three high-symmetry stackings, as listed in Table S1 of Supplementary Information.

The moiré band structures can be calculated by the plane-wave expansion. **Figs. 2(a-c)** plot moiré bands at three typical twist angles. One can clearly see that the Kramer pairs are formed at the time-reversal invariant momenta. This is because the moiré Hamiltonian Eq. (3) still preserves the time-reversal symmetry, as required by the type-II Rashba structure. However, the moiré bands are now lifted at the moiré Brillouin zone corner, indicating that the twisted BiTeI homobilayer breaks the inversion symmetry for the moiré superlattices. This is because these AA, AB, and AC stackings have different electron affinities, as shown in Supplementary Information, and are reflected by the nonzero value for $\phi$ in the moiré potential $\Delta(\mathbf{r})$. Moreover, within a twist angle $\theta$ such as 3.5°, isolated flat bands are observed ($C_1$-$C_4$) with the bandwidth about 7 meV. This is substantially narrower than those calculated in twisted TMDs. For example, it is around 20 or 40 meV for twisted $MoTe_2$ or $WSe_2$. [45] **Fig. 2(d)** summarizes the evolution of the bandwidth with the twist angle. Such an exceptional and tunable flatness indicates that twisted type-II Rashba structures can be a promising platform to realize and control many-electron correlation effects.

*Topological phase transition:* Then we study the topological properties of these isolated flat ($C_1$-$C_4$) bands. Notice that as the twist angle increases, valleys are formed at the +K/-K point of the moiré Brillouin zone as presented in **Figs. 2(a-c)**. Moreover, **Figs. 2 (e)** and **(f)** show that significant Berry curvature with equal values but opposite signs is distributed at the +K and -K points. Our berry curvature calculation further quantifies that the valley Chern number for the $C_1$ and $C_2$ ($C_3$ and $C_4$) bands is +1 (-1), featuring the valley Hall (VH) insulating states for twisted type-II Rashba BiTeI bilayers, which have been predicted and observed in non-twisted TMDs. [46,47]

Importantly, we find a critical twist angle 3.3° where the band gap between the $C_2$ and $C_3$ bands closes and form a gapless Dirac cone. The Berry curvatures for the $C_2$ and $C_3$ bands reverse their signs, as shown in **Figs. 2(e-f)**, suggesting a band inversion and topological phase transition. Because the twisted bilayer type-II Rashba BiTeI still preserves the time-reversal symmetry, its topology is characterized by the topological invariant $Z_2$, which can be calculated by the Wilson loop method that tracks the evolution of the Wannier charge center (WCC) plotted in the insets of **Fig. 2(b)** and **(c)**. As the twist angle becomes larger than the critical angle, the $Z_2$ topological invariant switches from 0 to 1 for the $C_1$ and $C_2$ bands, yielding a nontrivial topological insulating state when the $C_1$ and $C_2$ bands are filled. [48,49]

Next, we have conducted an in-depth study to understand the reasons behind this topological transition. Firstly, we find that in the VH insulating phase, the real-space wave functions of $C_1$ and $C_2$ bands at the +K/-K point are localized around the AC sublattices, while those of $C_3$ and $C_4$ bands are mostly localized around the AB sublattices (see Fig. S2 in Supplementary Information). Such a sublattice energy splitting contributes to a term proportional to $\sigma_z$, where $\sigma_z$ is a Pauli matrix that denotes the on-site energy difference of AC and AB sublattices.

Secondly, the expectation value of the spin operator, namely, $\langle\psi_n(\boldsymbol{k})|s_z|\psi_n(\boldsymbol{k})\rangle$ exhibits the opposite sign for each set of bands, as illustrated in **Fig. 3(a)**, while the expectation value of $s_{x,y}$ is almost vanishing over the moiré Brillouin zone, indicating the spin texture of the low energy moiré bands are along the out-of-plane direction. Our simulation shows that, at the +K point, spins are ↓/↑ for the $C_1$/$C_2$ band located at the AC sublattice, and ↑/↓ for the $C_3$/$C_4$ band located at the AB sublattice. Such a spin splitting indicates that the spins are coupled with sublattices by a term $\sigma_z s_z$. Moreover, at the -K point, spins for each band are reversed, or namely, ↑/↓ for the $C_1$/$C_2$ band at the AC sublattice, and ↓/↑ for the $C_3$/$C_4$ band at the AB sublattice, which can be described by a $-\sigma_z s_z$ term. In this sense, these states near the band gap can be captured by a phenomenological model, [50,51]

$$H = \Delta\sigma_z + \lambda\tau_z\sigma_z s_z, \qquad (4)$$

where $\tau_z$ denotes the valley index. This model preserves the time-reversal symmetry but breaks the inversion symmetry due to the first term. The second term is indeed the effective intrinsic SOC term that features the Kane-Mele model, and it results in the QSH insulating states of the twisted type-II Rashba systems.

Finally, we reveal that how the terms in the Kane-Mele model in Eq. (4) are harbored in the moiré Hamiltonian of Eq. (3). Because of flat bandwidth, we can neglect the kinetic energy term and reduce Eq. (3) to $H_{Moiré}(\boldsymbol{r},\boldsymbol{k}) = \Delta(\boldsymbol{r}) + \zeta_z\big(\alpha_R(k_x s_y - k_y s_x) + \alpha_I(\boldsymbol{r})s_z\big)$. Furthermore, around the critical bandgap closing at the $\pm K$ point, the spin texture is along the out-of-plane (z) direction, and the Rashba term approaches to zero. Thus, only two terms of the moiré Hamiltonian are important, $\Delta(\boldsymbol{r}) + \zeta_z\alpha_I(\boldsymbol{r})s_z$. Note that the scalar moiré potential $\Delta(\boldsymbol{r})$ of the twisted type-II Rashba BiTeI bilayer has two potential minimums around AB and AC stackings, which are about 100 meV lower than that of the AA stacking, as shown in **Figs. 3(b)** and **(c)**. In this scenario, this scalar moiré potential gives rise to Dirac-like bands at the moiré Brillouin zone corner as well as a honeycomb lattice with AB and AC being two sublattices, which is illustrated in **Fig. 3(b)**. Moreover, there is a tiny energy difference between AB and AC sublattices about 13 meV, as shown in **Fig. 3(c)**. Such a scalar moiré potential $\Delta(r)$ opens a trivial moiré band gap at the moiré Brillouin zone corner and results in a $\Delta\sigma_z$ term. Similarly, the interlayer interaction $\alpha_I(r)$ can also form a moiré band gap by with an extra prefactor, which is characterized by $\zeta_z\sigma_z$. Thus, the moiré Hamiltonian around the $\pm K$ point is essentially $\sigma_z + \zeta_z\sigma_z s_z$.

Notably, the layer pseudospins $\zeta_z$ in the Hamiltonian Eq. (3) couple with both momentum and $s_z$, when a gap is open at $k$, a gap with an opposite sign has to open at $-k$. Hence the layer pseudospins $\zeta_z$ play a similar role with the valley pseudospins $\tau_z$. There, the interlayer interaction $\alpha_I(r)$ works as an effective SOC. In this sense, the moiré Hamiltonian Eq. (3) essentially harbors the Kane-Mele model (Eq. (4)).

Generally, the effective SOC interaction is dependent on the gradient of the potential. the existence of effective SOC has also been demonstrated in other moiré systems, and it was revealed to be dependent on the gradient of moiré potential. [52,53] As the twist angle increases, the moiré period decreases, and the gradient of moiré potential increases. Consequently, the effective SOC strength dominates the gap over the sublattice energy difference for larger twist angles, driving the system into the QSH state.

Finally, the phase diagrams according to the Rashba effect and interlayer interaction are plotted. In **Fig. 4(a)**, we show the topological phases between the $C_2$ and $C_3$ bands as a function of the twist angle ($\theta$) and Rashba strength ($\alpha_R$). We restricted the angles ranging from 1 degree to 8 degrees. It is evident that VH and QSH states exists with a significant parameter range. This demonstrates that the observed topological phase transition as increasing twist angle is not incidental but rather intrinsic and robust. Moreover, when the Rashba strength reduces to about 0.3, the system cannot achieve a topological state within our considered angel range, indicating the importance of the Rashba effect for the form of topological states.

**Figure 4(b)** presents the coordinative contributions from interlayer interaction and Rashba strength to the insulating QSH state. As mentioned at the beginning, a challenge to realize topological states is to isolate the nontrivial band gap from the surrounded metallic states in Rashba materials. As seen from **Figs. 2 (a)-(c)**, the band energy at the K point needs to be higher than that at the $\Gamma$ point to result in a finite gap. **Fig. 4 (b)** plots this energy difference and shows that larger interlayer interaction $\alpha_I$ will lift the K-point energy and favor insulating states. This also agrees with **Fig. 1(b)**, in which the larger $\alpha_I$ lowers the band energy at the $\Gamma$ point. Finally, we observe that $\alpha_R$ and $\alpha_I$ increase linearly when the topological phase transition is satisfied, indicating that these two parameters need to be synchronized.

*Overlook:* We have established the mechanism wherein inversion-stacked Rashba materials can undergo topological phase transitions under twisting. The essential ingredient is a type-II Rashba moiré systems described at low energy by the universal Hamiltonian, Eq. (1). This theoretical framework can potentially be realized using a wide range of materials as consistent layers. Potential candidates include polar structures (e.g. $BiTeCl$, $Sb_2TeSe_2$, $TiNCl$), IV-VI materials (e.g. $GeTe$, $SnTe$, $PbS$), 2D ferroelectric materials (e.g. $In_2Se_3$), even some non-vdW Rashba materials (e.g. $BaCdK_2Sb_2$, $KSnAs$) via interface engineering. These classic Rashba materials are expected to undergo topological phase transitions driven by appropriate interlayer interactions and moiré potential, and they can be easily manipulated by external conditions such as stress or electric fields. Furthermore, the flatness of the energy bands induced by the Rashba effect makes them promising candidates for further realizing the integer or fractional QAH effect.

**Acknowledge:**

X.X. is supported by National Science Foundation (NSF) Designing Materials to Revolutionize and Engineer our Future (DMREF) DMR-2118779. H.W. and L.Y. are supported by NSF DMR-2124934. The simulation used Anvil at Purdue University through allocation DMR100005 from the Advanced Cyberinfrastructure Coordination Ecosystem: Services & Support (ACCESS) program, which is supported by National Science Foundation grants #2138259, #2138286, #2138307, #2137603, and #2138296.

**References:**

[1]  D. Leykam, A. Andreanov, and S. Flach, *Artificial Flat Band Systems: From Lattice Models to Experiments*, Advances in Physics: X **3**, 1473052 (2018).

[2]  Z. Jiang et al., *Flat Bands, Non-Trivial Band Topology and Rotation Symmetry Breaking in Layered Kagome-Lattice RbTi$_3$Bi$_5$*, Nat. Commun. **14**, 4892 (2023).

[3]  Y. Hu et al., *Non-Trivial Band Topology and Orbital-Selective Electronic Nematicity in a Titanium-Based Kagome Superconductor*, Nat. Phys. **19**, 1827 (2023).

[4]  Z. Zhang, Y. Wang, K. Watanabe, T. Taniguchi, K. Ueno, E. Tutuc, and B. J. LeRoy,

*Flat Bands in Twisted Bilayer Transition Metal Dichalcogenides*, Nat. Phys. **16**, 1093 (2020).

[5] L. Balents, C. R. Dean, D. K. Efetov, and A. F. Young, *Superconductivity and Strong Correlations in Moiré Flat Bands*, Nat. Phys. **16**, 725 (2020).

[6] J. Mao et al., *Evidence of Flat Bands and Correlated States in Buckled Graphene Superlattices*, Nature **584**, 215 (2020).

[7] M. Levin and A. Stern, *Fractional Topological Insulators*, Phys. Rev. Lett. **103**, 196803 (2009).

[8] D. N. Sheng, Z.-C. Gu, K. Sun, and L. Sheng, *Fractional Quantum Hall Effect in the Absence of Landau Levels*, Nat. Commun. **2**, 389 (2011).

[9] T. Neupert, L. Santos, C. Chamon, and C. Mudry, *Fractional Quantum Hall States at Zero Magnetic Field*, Phys. Rev. Lett. **106**, 236804 (2011).

[10] N. Regnault and B. A. Bernevig, *Fractional Chern Insulator*, Phys. Rev. X **1**, 021014 (2011).

[11] E. Tang, J.-W. Mei, and X.-G. Wen, *High-Temperature Fractional Quantum Hall States*, Phys. Rev. Lett. **106**, 236802 (2011).

[12] C. Wang, L. Gioia, and A. A. Burkov, *Fractional Quantum Hall Effect in Weyl Semimetals*, Phys. Rev. Lett. **124**, 096603 (2020).

[13] E. M. Spanton, A. A. Zibrov, H. Zhou, T. Taniguchi, K. Watanabe, M. P. Zaletel, and A. F. Young, *Observation of Fractional Chern Insulators in a van Der Waals Heterostructure*, Science **360**, 62 (2018).

[14] H. Park et al., *Observation of Fractionally Quantized Anomalous Hall Effect*, Nature **622**, 74 (2023).

[15] J. Cai et al., *Signatures of Fractional Quantum Anomalous Hall States in Twisted MoTe$_2$*, Nature **622**, 63 (2023).

[16] Z. Lu, T. Han, Y. Yao, A. P. Reddy, J. Yang, J. Seo, K. Watanabe, T. Taniguchi, L. Fu, and L. Ju, *Fractional Quantum Anomalous Hall Effect in Multilayer Graphene*, Nature **626**, 759 (2024).

[17] F. Xu et al., *Observation of Integer and Fractional Quantum Anomalous Hall Effects in Twisted Bilayer MoTe$_2$*, Phys. Rev. X **13**, 031037 (2023).

[18] C. Wang, X.-W. Zhang, X. Liu, Y. He, X. Xu, Y. Ran, T. Cao, and D. Xiao, *Fractional Chern Insulator in Twisted Bilayer MoTe$_2$*, Phys. Rev. Lett. **132**, 036501 (2024).

[19] X. Liu, Y. He, C. Wang, X.-W. Zhang, T. Cao, and D. Xiao, *Gate-Tunable Antiferromagnetic Chern Insulator in Twisted Bilayer Transition Metal Dichalcogenides*, Phys. Rev. Lett. **132**, 146401 (2024).

[20] Y. Zhang, T. Devakul, and L. Fu, *Spin-Textured Chern Bands in AB-Stacked Transition Metal Dichalcogenide Bilayers*, Proc. Natl. Acad. Sci. U.S.A. **118**, e2112673118 (2021).

[21] T. Devakul, V. Crépel, Y. Zhang, and L. Fu, *Magic in Twisted Transition Metal Dichalcogenide Bilayers*, Nat. Commun. **12**, 6730 (2021).

[22] M. Claassen, L. Xian, D. M. Kennes, and A. Rubio, *Ultra-Strong Spin–Orbit Coupling and Topological Moiré Engineering in Twisted ZrS2 Bilayers*, Nat. Commun. **13**, 4915 (2022).

[23] F. Wu, T. Lovorn, E. Tutuc, and A. H. MacDonald, *Hubbard Model Physics in Transition Metal Dichalcogenide Moiré Bands*, Phys. Rev. Lett. **121**, 026402 (2018).

[24] F. Wu, T. Lovorn, E. Tutuc, I. Martin, and A. H. MacDonald, *Topological Insulators in Twisted Transition Metal Dichalcogenide Homobilayers*, Phys. Rev. Lett. **122**, 086402 (2019).

[25] G. Bihlmayer, P. Noël, D. V. Vyalikh, E. V. Chulkov, and A. Manchon, *Rashba-like Physics in Condensed Matter*, Nat. Rev. Phys. **4**, 642 (2022).

[26] Y. A. Bychkov and E. I. Rashba, *Oscillatory Effects and the Magnetic Susceptibility of Carriers in Inversion Layers*, Journal of Physics C: Solid State Physics **17**, 6039 (1984).

[27] S. Lee, H. Koike, M. Goto, S. Miwa, Y. Suzuki, N. Yamashita, R. Ohshima, E. Shigematsu, Y. Ando, and M. Shiraishi, *Synthetic Rashba Spin–Orbit System Using a Silicon Metal-Oxide Semiconductor*, Nat. Mater. **20**, 1228 (2021).

[28] K. Miyata and X. Y. Zhu, *Ferroelectric Large Polarons*, Nat. Mater. **17**, 379 (2018).

[29] S. D. Ganichev et al., *Experimental Separation of Rashba and Dresselhaus Spin Splittings in Semiconductor Quantum Wells*, Phys. Rev. Lett. **92**, 256601 (2004).

[30] H. Nakamura, T. Koga, and T. Kimura, *Experimental Evidence of Cubic Rashba*

*Effect in an Inversion-Symmetric Oxide*, Phys. Rev. Lett. **108**, 206601 (2012).

[31] G. Bihlmayer, O. Rader, and R. Winkler, *Focus on the Rashba Effect*, New J. Phys. **17**, 050202 (2015).

[32] X. Zhang, Q. Liu, J. W. Luo, A. J. Freeman, and A. Zunger, *Hidden Spin Polarization in Inversion-Symmetric Bulk Crystals*, Nat. Phys. **10**, 387 (2014).

[33] S. Lee, M. Kim, and Y. K. Kwon, *Unconventional Hidden Rashba Effects in Two-Dimensional InTe*, NPJ 2D Mater. Appl. **7**, 43 (2023).

[34] J. H. Ryoo and C.-H. Park, *Hidden Orbital Polarization in Diamond, Silicon, Germanium, Gallium Arsenide and Layered Materials*, NPG Asia Mater. **9**, e382 (2017).

[35] G. Gatti et al., *Hidden Bulk and Surface Effects in the Spin Polarization of the Nodal-Line Semimetal ZrSiTe*, Commun. Phys. **4**, 54 (2021).

[36] L. Yuan, Q. Liu, X. Zhang, J. W. Luo, S. S. Li, and A. Zunger, *Uncovering and Tailoring Hidden Rashba Spin–Orbit Splitting in Centrosymmetric Crystals*, Nat. Commun. **10**, 906 (2019).

[37] M. S. Bahramy, R. Arita, and N. Nagaosa, *Origin of Giant Bulk Rashba Splitting: Application to BiTeI*, Phys. Rev. B **84**, 041202 (2011).

[38] K. Ishizaka et al., *Giant Rashba-Type Spin Splitting in Bulk BiTeI*, Nat. Mater. **10**, 521 (2011).

[39] A. Crepaldi et al., *Giant Ambipolar Rashba Effect in the Semiconductor BiTeI*, Phys. Rev. Lett. **109**, 096803 (2012).

[40] G. Bianca et al., *Liquid-Phase Exfoliation of Bismuth Telluride Iodide (BiTeI): Structural and Optical Properties of Single-/Few-Layer Flakes*, ACS Appl. Mater. Interfaces, **14**, 34963 (2022).

[41] N. Antonatos, E. Kovalska, V. Mazánek, M. Veselý, D. Sedmidubský, B. Wu, and Z. Sofer, *Electrochemical Exfoliation of Janus-like BiTeI Nanosheets for Electrocatalytic Nitrogen Reduction*, ACS Appl. Nano Mater. **4**, 590 (2021).

[42] H. Maaß et al., *Spin-Texture Inversion in the Giant Rashba Semiconductor BiTeI*, Nat. Commun. **7**, 11621 (2016).

[43] J. S. Lee, G. A. H. Schober, M. S. Bahramy, H. Murakawa, Y. Onose, R. Arita, N. Nagaosa, and Y. Tokura, *Optical Response of Relativistic Electrons in the Polar BiTeI*

*Semiconductor*, Phys. Rev. Lett. **107**, 117401 (2011).

[44] C. Mera Acosta, E. Ogoshi, A. Fazzio, G. M. Dalpian, and A. Zunger, *The Rashba Scale: Emergence of Band Anti-Crossing as a Design Principle for Materials with Large Rashba Coefficient*, Matter **3**, 145 (2020).

[45] G. Yu, L. Wen, G. Luo, and Y. Wang, *Band Structures and Topological Properties of Twisted Bilayer $MoTe_2$ and $WSe_2$*, Phys. Scr. **96**, 125874 (2021).

[46] H. Rostami, F. Guinea, M. Polini, and R. Roldán, *Piezoelectricity and Valley Chern Number in Inhomogeneous Hexagonal 2D Crystals*, NPJ 2D Mater. Appl. **2**, 15 (2018).

[47] F. Zhang, A. H. MacDonald, and E. J. Mele, *Valley Chern Numbers and Boundary Modes in Gapped Bilayer Graphene*, Proc. Natl. Acad. Sci. U.S.A. **110**, 10546 (2013).

[48] W.-X. Qiu, B. Li, X.-J. Luo, and F. Wu, *Interaction-Driven Topological Phase Diagram of Twisted Bilayer $MoTe_2$*, Phys. Rev. X **13**, 041026 (2023).

[49] X.-W. Zhang, C. Wang, X. Liu, Y. Fan, T. Cao, and D. Xiao, *Polarization-Driven Band Topology Evolution in Twisted $MoTe_2$ and $WSe_2$*, Nat. Commun. **15**, 4223 (2024).

[50] C. L. Kane and E. J. Mele, *$Z_2$ Topological Order and the Quantum Spin Hall Effect*, Phys. Rev. Lett. **95**, 146802 (2005).

[51] C. L. Kane and E. J. Mele, *Quantum Spin Hall Effect in Graphene*, Phys. Rev. Lett. **95**, 226801 (2005).

[52] Z. Liu, H. Wang, and J. Wang, *Magnetic Moiré Surface States and Flat Chern Bands in Topological Insulators*, Phys. Rev. B **106**, 035114 (2022).

[53] H. Wang and L. Yang, *Topological Phase Transition from Periodic Edge States in Moiré Superlattices*, Phys. Rev. B **107**, 235427 (2023).

**Figures:**

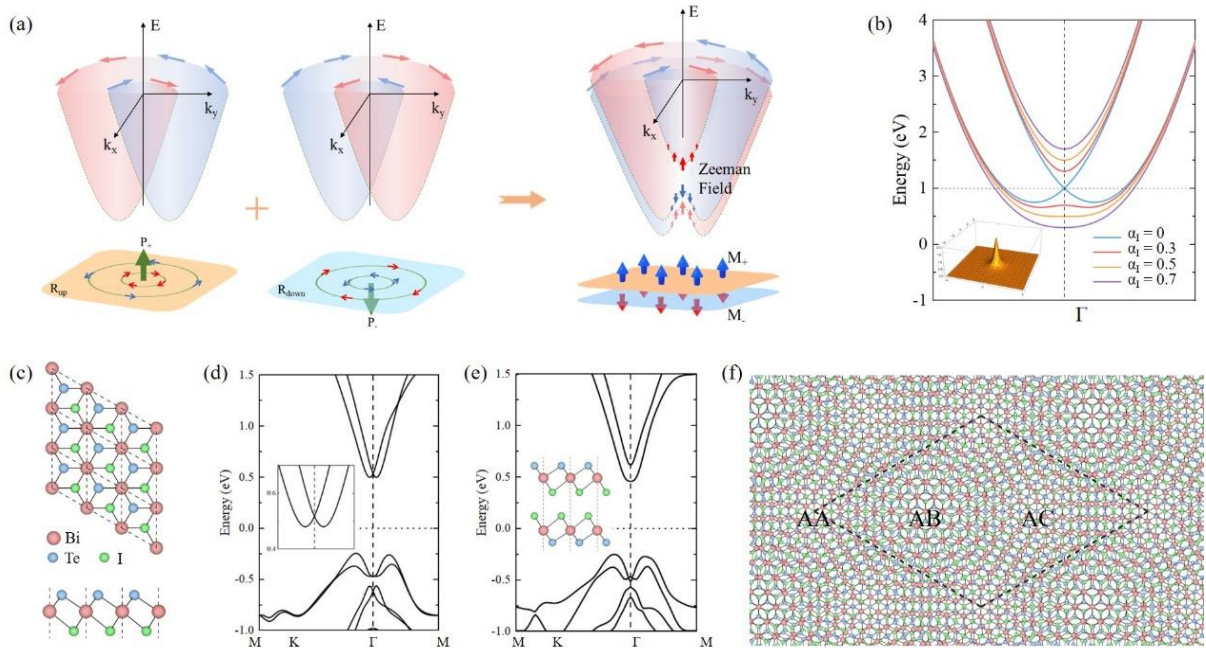

**Figure 1**. (a) Schematic of type-II Rashba band structure and spin textures. (b) Bands of a type-II Rashba bilayer with different interlayer couplings. The inset is the calculated berry curvature of the lowest-energy band around the Γ point. (c) Crystal structures of monolayer BiTeI. (d) and (e) Band structures of monolayer BiTeI and that of type-II Rashba bilayer. (f) Top view of twisted bilayer BiTeI.

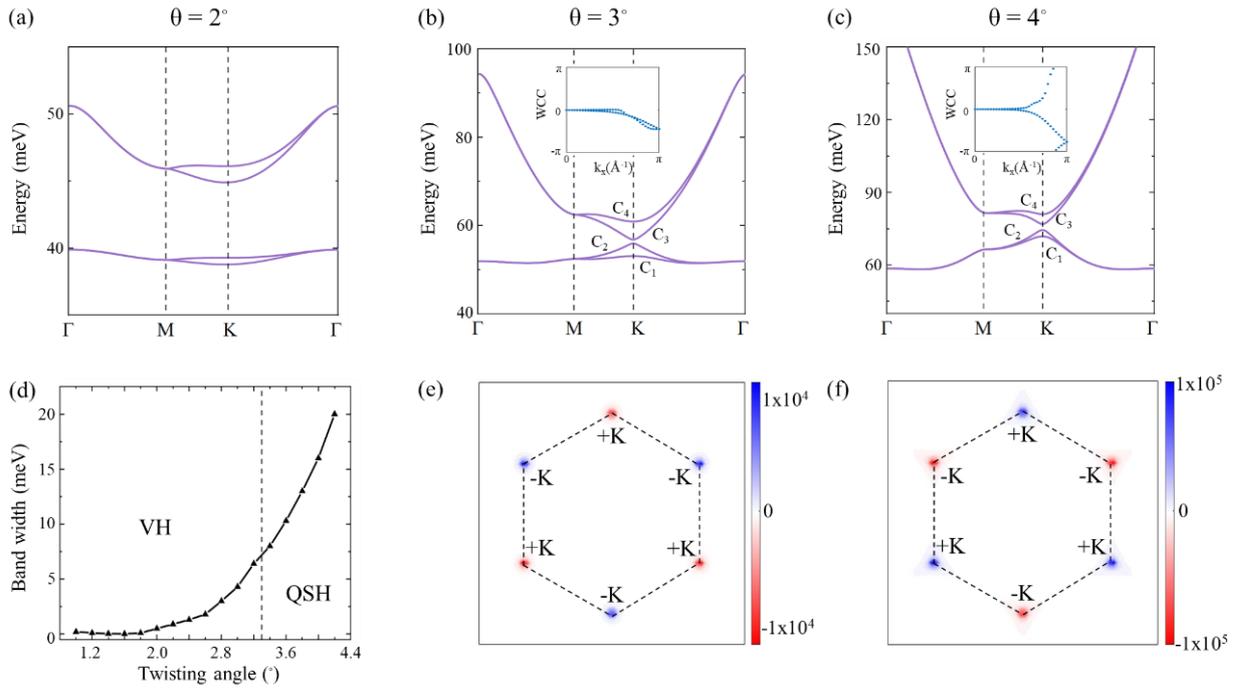

**Figure 2** Moiré bands with a twisting angle of 2° (a), 3° (b) and 4° (c). The insets in (b) and (c) illustrate the evolution of Wannier charge center. (d) Variation of the band width as a function of the twisting angle. (e) and (f) Berry curvature of the two lowest conduction bands with a twisting angle of 3° and 4°, respectively.

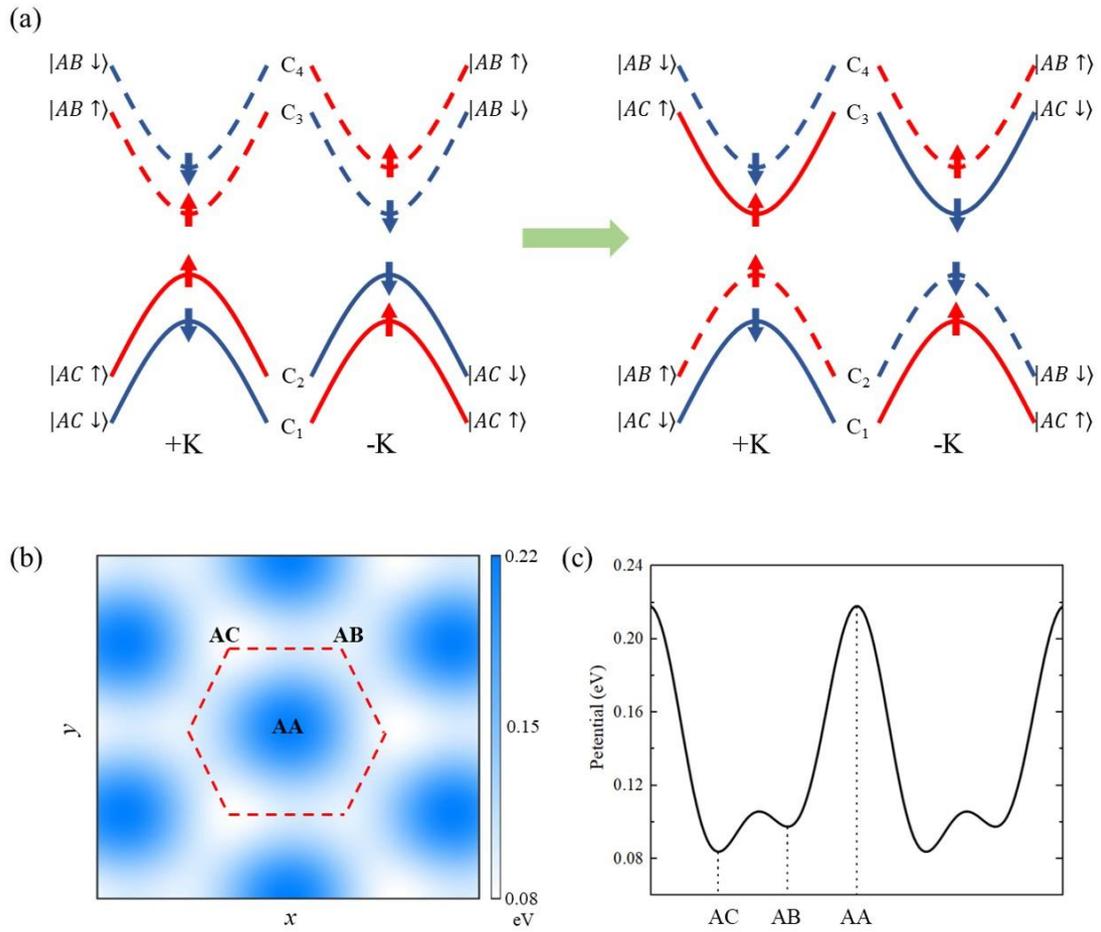

**Figure 3** (a) Schematic plots of bands before and after the topological phase transition. The arrow on each band denotes the spin orientation. (b) and (c) Top and side views of the moiré potential of twisting bilayer BiTeI, respectively.

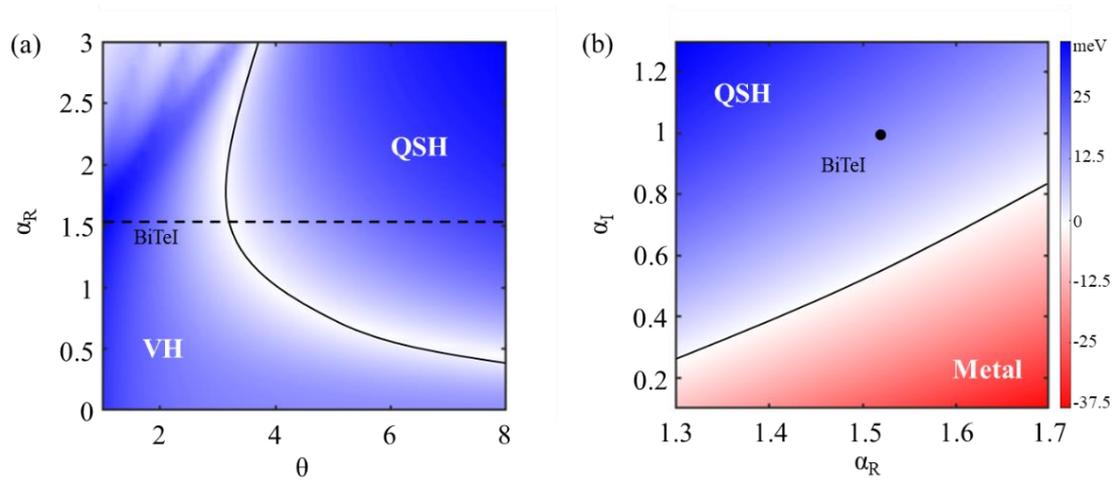

**Figure 4** (a) Phase diagram with the twist angle and Rashba parameter. (b) That with the strength of Rashba parameter and interlayer interaction. The dashed line represents the parameter used in twist bilayer BiTeI. The solid line represents the critical phase transition boundary.